\documentstyle[12pt]{article}
\begin{document}
\begin{center}
{\large \bf 
Dynamical Behavior of Continuous Tick Data in Futures Exchange Market
}\\

\vspace*{.5in}

\normalsize 
Kyungsik Kim$^{*}$ and Seong-Min Yoon \\

\vspace*{.2in}

{\em 
$^{*}$$Department$ $of$ $Physics$, $Pukyong$ $National$ $University$,\\
$Pusan$ $608$-$737$, $Korea$\\
$Division$ $of$ $Economics$, $Pukyong$ $National$ $University$,\\
$Pusan$ $608$-$737$, $Korea$ \\

}

\hfill\\
\end{center} 
%
 
%
%
%
\baselineskip 24pt
\begin{center}
{\bf Abstract}
\end{center} 

\noindent
We study the tick dynamical behavior of the bond futures in Korean Futures Exchange(KOFEX) market.
Since the survival probability in the continuous-time random walk theory is applied to the bond futures transaction,
the form of the decay function in our bond futures model is discussed from two kinds of Korean Treasury Bond(KTB) 
transacted recently in KOFEX.
The decay distributions for survival probability are particularly displayed stretched exponential forms
with novel scaling exponents $\beta$ $=$ $0.82$(KTB $203$) and $\beta$ $=$ $0.90$(KTB$112$), 
respectively, for our small time intervals. 
We obtain the scaling exponents for survival probability 
$\epsilon$ $=$ $17$ and $18$ decayed rapidly in large time limit,
and our results are compared with recent numerical calculations.  

\vskip 35mm
\noindent
$Keywords$: Continuous-time random walk; Bond futures; Survival probability; Stretched exponential function                                                         
\vskip 5mm
\noindent
$^{*}$$E-mail$ $address$: kskim@pknu.ac.kr.

\newpage

\noindent
{\bf 1. INTRODUCTION}
\hfill\\


Recently, the investigation of economic phenomena between economists and physicists
has received considerable attention as one interdisciplinary field.$^{1,2}$ 
Among many outstanding topics, there has been mainly concentrated on 
price changes in open market,$^{3,4}$ the distribution of income of companies,
the scaling relation of company's size fluctuations, the financial analysis of foreign 
exchange rates,$^{5-7}$ and tick data analysis of bond futures.$^{8-10}$ 
Particularly, the study of fluctuations in economics can lead to a better understanding for
the scaling properties based on methods and approaches of physics.
The power law distribution, the stretched exponential
distribution, and L$\acute{e}$vy-like distribution have been really elucidated the general mechanism
in diverse fields of science and technology. Mantegna and Stanley$^{4}$ reported that price
changes in a stock market scales as a power law of the index change with a scaling exponent $\alpha=1.4$.   
Takayasu $et$ $al.$$^{5}$ have showed the stretched exponential distribution from 
results of analysis observed by tick data of yen-dollar exchange rates
and have recently investigated extensively several hyper-inflations$^{11}$ in which the rapid growth of price changes
is approximated by a double exponential function.         
Lee $at$ $al.$$^{12}$ have obtained the fluctuation in the gross domestic product that
the width of the distribution scales as a power law of
the gross domestic product with a scaling exponent $\beta=0.15$.
Otherwise, many works$^{5,6,7,13,14}$ have reported that the price fluctuations follow the anomalous power law
from the stochastic time evolution equation, i.e. Langevin-type equation. 

In the stochastic process, one of the well-known popular subjects for the
regular and disordered systems is the random walk theory,$^{15}$ and this theory has 
extensively developed to the continuous-time random walk theory which is
essentially described both by the transition probability dependent of the 
length between steps and by the distribution of the pausing times.$^{16}$
Until now the transport phenomena for the motion of tracers
have largely been extended to the reaction kinetics,$^{17,18}$ the strange kinetics,$^{19}$
fractional diffusion equations,$^{20,21}$ and generalized Langevin and Fokker-Planck equations.$^{22-26}$     

Recently, Scalas$^{8}$ presented the correlation function for bond walks from the time series of 
BUND and BTP(Buoni del tesoro Poliennali) futures exchanged at 
the London International Financial Futures and options Exchange(LIFFE).  
Scalas $et$ $al.$$^{9}$ have discussed that the continuous-time random walk theory,
formerly introduced by Montroll and Weiss,$^{16}$, is applied to the dynamical behavior
by tick-by-tick data in financial markets.
Mainardi $et$ $al.$$^{10}$ have also observed the waiting-time distribution for bond futures traded at 
LIFFE, London.
In this paper, theoretical and numerical arguments 
for the volume of bond futures traded at KOFEX of Pusan are presented.
Mainly, we consider the number of transactions for two different delivery dates and find
the decay functions for survival probabilities in our bond futures model.
Sec. 2 is devoted to the survival probability in the continuous-time random walk theory and
to the decay function of survival probabilities. In Sec. 3, we reports numerical results and some 
concluding remarks.\\

\noindent
{\bf 2. THE SURVIVAL PROBABILITY AND THE STRETCHED EXPONENTIAL FUNCTION}
\hfill\\

First of all, we will focus on the continuous-time random walk theory$^{16}$ 
in order to discuss the survival probability. 
Since the pausing-time distribution $ \psi(t)$ between two consecutive 
transactions and the price $l(t)$ can be modelled as random variables in financial markets,
we consider that the coupled transition probability density $\Psi(l,t)$ for jump $l$ and pausing-time 
$t$ is defined by
\begin{equation}
\Psi(l,t)=p(l)\psi(t)
\label{eq:a1}
\end{equation}
where $p(l)$ is the transition probability dependent of the 
length between steps and $ \psi(t)$ is the pausing-time distribution.  
Summing over all $l$ with the periodic boundary condition,
Eq.$(1)$ is given by
\begin{equation}
\sum_{l} \Psi(l,t) =\sum_{l} p(l)\psi(t)=\psi(t).
\label{eq:b2}
\end{equation}
Since $R_n (l,t)$ is the probability density arriving immediately at lattice point $l$ at time
$t$ after one random walker goes to $n$ steps, $R_n (l,t)$ is satisfied with the recursion relation as
\begin{equation}
R_n (l,t)= \sum_{l'} \int_{0}^{t} dt' p(l-l') \psi(t-t') R_{n-1} (l',t').
\label{eq:c3}
\end{equation}
In order to find the probability density, $P(l,t)$, which is existed at lattice point $l$ at time
$t$, we introduce the relation between $P(l,t)$ and $R (l,t-t')$ such as
\begin{equation}
P (l,t)=  \int_{0}^{t} dt' \psi(t') R(l,t-t'),
\label{eq:e5}
\end{equation} 
where
\begin{equation}
R (l,t)=  \sum_{n=0}^{\infty} \psi_n (t) P (l,n).
\label{eq:d4}
\end{equation}

The survival probability, $\Psi(t)$, which is stayed for the time $t$ after arrivng 
at an arbitrary lattice point, can be expressed in terms of
\begin{equation}
\Psi(t) = 1-  \int_{0}^{t} dt' \sum_{l} \Psi(l,t')  
        = 1-  \int_{0}^{t} dt' \psi(t').  
\label{eq:f6}
\end{equation} 
Hence, after substituting Eq. $(6)$ into Eq. $(5)$, the Fourier-Laplace transform of the probability
density, i.e. Montroll-Weiss equation, is described$^{16}$ as
\begin{eqnarray}
P(q,u)  &=&   \sum_{l}  \int_{0}^{\infty} dt P(l,t) \exp (-iql-ut)  \nonumber \\
       &=& \frac{R(q,u)[1-\psi(u)]}{u}  \nonumber \\
       &=& \frac{P(q,0)}{[1-\Psi(q,u)]} \frac{[1-\psi(u)]}{u}, 
\label{eq:g7}
\end{eqnarray} 
where the generalized structure function, i.e.  the Fourier-Laplace transform of Eq. $(1)$ is
given by
\begin{equation}
\Psi(q,u)= p(q)\psi(u)
\label{g7}
\end{equation} 
and $P(q,0)$ is the Fourier transform of the initial condition $P(l,0)$.

Let us introduce the following scaling behavior in the long jump and long time limits: 
\begin{equation}
p(q) \sim 1- \mid q \mid^{\alpha},  q \rightarrow 0,  0 < \alpha \leq 2,  
\label{eq:bb22}
\end{equation} 
and
\begin{equation}
\psi(u) \sim 1- u^{\beta},  u \rightarrow 0,  0 < \beta \leq 1.   
\label{eq:cc23}
\end{equation} 
We can also present approximately as 
\begin{equation}
p(q) = \exp (- \mid q \mid^{\alpha}), 0 < \alpha \leq 2  
\label{eq:dd24}
\end{equation} 
and
\begin{equation}
\psi(u) = \frac{1}{ 1+ u^{\beta}}, 0 < \beta \leq 1,   
\label{eq:ee25}
\end{equation} 
where Eq. $(11)$ means one type of the characteristic function for L$\acute{e}$vy stable distribution.
Hence the inverse Fourier transform of Eq. $(11)$ is found to scale
approximately with $p(l)$ $\simeq$ $ \mid l \mid^{-(\alpha+1)}$ for $ 0 < \alpha < 2 $
in the large  $ l $ limit.     

Since the Laplace transform of survival probability from Eq. $(6)$ is given by 
\begin{equation}
\Psi(u) = \frac{1-\psi(u)}{u},    
\label{eq:aa21}
\end{equation} 
we find that 
\begin{equation}
\Psi(u) = \frac{u^{\beta-1}}{1+u^{\beta}}, 
\label{eq:ff27}
\end{equation} 
after substituting  Eq. $(12)$ into Eq. $(13)$.
The survival probability, i.e. the inverse Laplace tranform of Eq. $(14)$, can be
derived as 
\begin{equation}
\Psi(t) = E_{\beta} (-t^{\beta}),  
\label{eq:gg28}
\end{equation} 
where the Mittag-Leffler function of order $\beta$$^{27,28}$ is
\begin{equation}
E_{\beta} (-t^{\beta}) =  \sum_{n=0}^{\infty} (-1)^n \frac{t^{\beta n}}{\Gamma(\beta n +1)}.  
\label{eq:hh29}
\end{equation} 
We also note that the decay distribution of the survival probability appears in the fractional 
relaxation process for $0 < \beta < 1$ while it includes to the fractional oscillation process
in the case of $1 < \beta < 2$. 

According to Eq. $(16)$, the survival probability $\Psi(t)$
for sufficiently small times reduces to a stretched exponential:
\begin{eqnarray}
E_{\beta} [-{(\gamma t)}^{\beta}] & \simeq & 1- \frac{{(\gamma t)}^{\beta n}}{\Gamma(\beta n +1)}  \nonumber \\ 
                       & \simeq & \exp [- \frac{{(\gamma t)}^{\beta}}{\Gamma(\beta +1)} ],  
\label{eq:jj31}
\end{eqnarray} 
where $\gamma$ is a time-scale factor. In the large $t$ limit, the survival probability $\Psi(t)$ from
remained transaction quantities less than $10$ percentages of total tick data scales as a power law 
\begin{equation}
\Psi(t) \sim - \epsilon \log t , 
\label{eq:kk32}
\end{equation} 
where $\epsilon$ is the scaling exponent for large times.
In our scheme we will make use of Eq. $(16)$-$(18)$ to find out the dynamical behavior
for the survival probability, and these mathematical functions would be expected to lead
us to more results.\\

\noindent
{\bf 3. RESULTS AND CONCUSIONS}
\hfill\\

In order to analyze survival probabilities for bond futures with delivery date,
our underlying asset is a Korean Government bond traded at KOFEX. 
Since the continuous-time random walk theory can sucessfully account for the time evolution 
of financial markets that is non-local and non-Markovian, 
we concentrate on the survival probabilities $\Psi(t)$ independent of transaction places.
We here consider two different delivery dates : December KTB$112$ and March KTB$203$. 
The tick data for December KTB$112$ were taken from July $2001$ to December $2001$
while we used the tick data of March KTB$203$ transacted for six months from October $2001$.
The KOFEX opens for $6$ hours per one day except for weekends, and the total tick data of transactions are,
respectively, about $2.3\times 10^5$ (KTB$112$) and $2.1\times 10^5$ (KTB$203$). 

Figs. $1$ and $2$ depict, respectively, the survival probability $\Psi(t)$ 
as a function of time step $t$ for December KTB$112$ and March KTB$203$ delivery dates. 
We will assume that one time step is the transaction time evoluted for 10 seconds, and total time
steps are approximately $2.9\times 10^5$ (KTB$112$) and $2.5\times 10^5$ (KTB$203$), 
neglecting time intervals of the market closure.
The slope of the solid line from Fig. $1$ gives approximately the streched exponential
function of Eq. $(17)$ with $\beta=0.90$ and $\gamma=1.5\times 10^{-5}$
for December KTB$112$ delivery date while the survival probability $\Psi(t)$ 
decays with $\beta=0.82$ and $\gamma=2.0\times 10^{-5}$ 
from Fig. $2$ of March KTB$203$ delivery date.  
Figs. $3$ and $4$ is also the plots of $\log$ $\Psi(t)$ as a function of $\log$ $t$ in Eq. $(18)$,
by taking continuous tick data as shown in Figs. $1$ and $2$. 
On the other hand, we can, respectively, obtain approximately the scaling exponents $\epsilon$ $=$ $17$(KTB$112$) 
and $18$(KTB$203$) decayed rapidly in large time $t$ limit, as plotted in the dashed line of Figs. $3$ and $4$.   

In conclusion, we have presented the dynamical behavior of the continuous tick data for our bond futures,
based on a continuous-time random walk theory recently proposed by Scalas $et$ $al.$$^{9,10}$
For two kinds of Korean Treasury Bond(KTB$203$ and KTB$112$) transacted recently in KOFEX, 
the decay distribution for survival probability in our small time intervals is displayed stretched exponential form 
while it scales as a power law for large times.
  
In Scalas model, they introduce BUND futures with two different delivery dates : June $1997$ and
September $1997$. The survival probabilities for their BUND futures have the form of
the well-known Mittag-Leffler function interpolated between two limitting behaviors: 
the stretched exponential function $ \exp [- {(\gamma t)}^{\beta} / \Gamma(\beta +1)] $ 
for small time intervals, and the power-law function ${(\gamma t)}^{-\beta}/{\Gamma(1- \beta )}$
for larger time intervals. The survival probability has an index $\beta=0.96$ and 
a time scaling factor $\gamma=1/12$ for BUND futures of June $1997$, in Figs.$1-4$ of Ref.$[10]$,
while it has $\beta=0.95$ and $\gamma=1/12$ for BUND futures of September $1997$.
As shown in Figs. $1$ and $2$, we have the index $\beta=0.90$ and a time scaling factor $\gamma=1.5\times 10^{-5}$
for KTB$112$ and $\beta=0.82$ and $\gamma=2.0\times 10^{-5}$ for KTB$203$ in our small time intervals.
However, our key results for the index $\beta$ arenot in good agreement with those of Scalas model.
It is in practice found that our scaling exponents $\beta$ are smaller than those of Scalas model$^{8,9}$,
because their transaction quantities are slightly less than ours. As transaction quantities increase more and more,
we would suggest that the scaling exponent $\beta$ take a value near $\frac{1}{2}$.
Hence, when $\beta$ is varied in the range $\frac{1}{2}$$\le$$\beta$$<$$\beta_{c}=1$,
it can be available for several future models to carry out the continuous-time random walk.
Here $\beta_{c}$ is the critical exponent which has the so-called marginal value of $\beta$ from
the dynamical behavior obtained in Scalas model.    
If transaction quantities decrease less than those of Scalas model, then the value $\beta$ becomes a larger value than one,
and for this case the curve of the survival probability may be decreased rapidly.      
 
In future, we will present the correlations for bond walk from the tick data of 
bond futures in Korean financial markets and in detail compare our results with bond futures 
transacted in other nations. 
It is also expected that the detail description of
the continuous-time random walk theory will be used to study 
extensions of foreign financial analysis of won(Korea)-yen(Japan) and won(Korea)-dollar(USA) exchange rates.\\

\vskip 3mm
\noindent
{\bf ACKNOWLEDGMENTS}
\vskip 2mm
\hfill\\
This work was supported by grant No. 2000-2-133300-001-3 from the Basic Research 
Program of the Korea Science and Engineering Foundation.

\vspace {5mm}
%

%
%
\newpage
%
\begin{center}
{\bf FIGURE  CAPTIONS}
\end{center}

\vspace {10mm}

\noindent
Figure $1$.  Plot of the survival probability for December KTB$112$ as a function of the time $t$.
The solid line is the stretched exponential function for $\beta=0.90$ and $\gamma=1.5\times 10^{-5}$. 
The continuous tick data for December KTB$112$ were taken from July $2001$ to December $2001$.
\vspace {15mm}

\noindent
Figure $2$.  Plot of the survival probability for March KTB$203$ as a function of the time $t$.
The solid line is the stretched exponential function for $\beta=0.82$ and $\gamma=2.0\times 10^{-5}$. 
The continuous tick data for March KTB$203$ were taken from October $2001$ to March $2002$.
\vspace {15mm}

\noindent
Figure $3$. Log-log plot of the survival probability of the time $t$ for December KTB$112$.
The solid line is the stretched exponential function for $\beta=0.90$ and $\gamma=1.5\times 10^{-5}$,
and the dashed line is scaled approximately as a power law with a scaling exponent $\epsilon = 17$.
\vspace {15mm}

\noindent
Figure $4$.  Log-log plot of the survival probability of the time $t$ for March KTB$203$.
The solid line is the stretched exponential function for $\beta=0.82$ and $\gamma=2.0\times 10^{-5}$,
and the dashed line is scaled approximately as a power law with a scaling exponent $\epsilon = 18$.

\end{document}